\documentstyle[aps,prl,multicol,fancyheadings,amsbsy,amssymb,amstex]{revtex}
\begin{document}
\title{Quantum-Mechanical Position Operator and Localization in Extended 
Systems}
\author{A. A. Aligia$^{a}$ and G. Ortiz$^{b}$}
\address{$^{a}$Comisi\'{o}n Nacional de Energ{\'{\i }}a At\'{o}mica, 
Centro At\'omico Bariloche and Instituto Balseiro, \\
8400 S.C. de Bariloche, Argentina}
\address{$^{b}$Theoretical Division, 
Los Alamos National Laboratory, Los Alamos, NM 87545}

\date{Received \today }

\maketitle

\begin{abstract}
We introduce a fundamental complex quantity, $z_{L}$, which allows us to
discriminate between a conducting and non-conducting thermodynamic phase in
extended quantum systems. Its phase can be related to the expectation value 
of the position operator, while its modulus provides an appropriate 
definition of a localization length.
The expressions are valid for {\it any} fractional particle filling. As an
illustration we use $z_{L}$ to characterize insulator to 
``superconducting'' and Mott transitions in one-dimensional lattice models
with infinite on-site Coulomb repulsion at quarter filling.
\end{abstract}

\vspace*{-1.1cm}
\begin{multicols}{2}

\columnseprule 0pt

\narrowtext

Macroscopically, the fundamental property that distinguishes an insulator
from a conductor is that a steady current cannot flow at zero temperature 
\cite{landau}. Understanding the quantum nature of the insulating state 
and possible phase transitions to a conducting state is a significant
endeavor which started with the seminal work of Kohn \cite{kohn}.
Localization of the electronic wave function and the
existence of a dielectric polarization density field are two related
features of the insulating ground state (GS). Accordingly, one would expect 
that wisdom developed from the recent microscopic theory of polarization 
\cite{ort} could be exploited to establish a criterion for localization.

In this Letter we introduce a complex quantity, $z_{L}$, which enables us to
distinguish between conducting (metal, superconductor) and non-conducting 
(band, Peierls, Anderson, and Mott insulators) states of matter. Its phase 
corresponds to the GS expectation value of the position operator, 
intrinsically connected to macroscopic polarization \cite{res1}, while its 
modulus provides an unambiguous definition of a localization length which 
can be described entirely in terms of the properties of the many-body ground 
state. In this way, $z_{L}$ generalizes the concept and quantity $z_{N}$ 
recently introduced by Resta and Sorella \cite{res2} which, as we will show 
below, is valid only at integer particle filling. Furthermore, we show the 
usefulness of these ideas to characterize metal-insulator or
superconducting-insulator quantum phase transitions in strongly correlated 
lattice models. 
 
To describe the intrinsic bulk properties of extended quantum systems it is 
almost mandatory to assume $N$ interacting 
particles enclosed in a box subjected to periodic Born-von K\'{a}rm\'{a}n 
(BvK) boundary conditions (BC), to reduce surface effects. On the other 
hand, the fact that its quantum state is defined in a non-simply connected 
manifold makes the center of mass of its wave function an ill-defined concept. 
In recent interesting papers \cite{res1,res2}, the appropriate definition of 
the GS expectation value of the position operator 
$\langle \hat{X}\rangle$ \cite{res1} and the 
localization length $\lambda$ \cite{res2} in one-dimensional (1D) systems 
satisfying BvK BC were discussed. The primary quantity was 
\begin{equation} 
z_{N}=\langle g|e^{i\frac{2\pi }{L}\hat{X}}|g\rangle \ ,  \label{f1} 
\end{equation} 
where $|g\rangle$ is the many-body GS of the system ($\langle 
g|g\rangle=1$), $\hat{X}=\sum_{j=1}^{N}x_{j}$ is the sum of particle positions 
($\hat{X}/N$ is the center of mass coordinate), and $L$ is 
the number of unit cells in the system.  $\langle \hat{X}\rangle$ 
was defined from the phase of $z_{N}$, 
while $\lambda$ was specified from the modulus of $z_{N}$, 
taking the thermodynamic limit ($N,L \rightarrow \infty$, keeping 
the density $n_{0}=N/L$ constant). 
The crux of these ideas is the expectation that, in the thermodynamic 
limit, $z_{N}$ can attain only two distinct values: $z_{N}$ vanishes for a 
conducting phase, while $|z_{N}|\rightarrow 1$ for an insulator. 
 
In microscopic 
calculations, translational invariance and BvK (or twisted) BC are 
convenient, and in most cases imperative, since the Hamiltonian 
can be diagonalized separately in each 
Hilbert subspace of the many-body states which belong to the same 
irreducible representation of the translation group (or space group, in 
general).  For systems which do not 
possess translational invariance, BvK (or twisted) BC are not an advantage 
over open BC, and for the latter, the traditional definition of the 
expectation value of the position operator (i.e. the first moment of the 
modulus squared of the many-body wave function, $|g|^{2}$, or ``dipole'') 
is a well-defined concept. Thus, in the following we concentrate only on 
translational invariant systems. 
 
For these systems, excluding accidental degeneracies, $|g\rangle$ belongs to 
a well-defined irreducible representation of the space group. In particular, 
if the GS is non degenerate (as assumed in Refs. \cite{res1,res2}), 
calling $\hat{T}$ the operator which translates one unit cell (taken as 
the unit of length) to the right, one has $\hat{T} \ |g\rangle = e^{iK} \ 
|g\rangle $, where $K$ is the total momentum of the GS. 
Degeneracies of states which differ in $K$ or any other conserved quantum 
number do not affect the arguments given below because the Hamiltonian does 
not mix them. In other words, energy level crossings are allowed whenever 
the states involved belong to different symmetry sectors. 
 
In order for the matrix element in Eq. (\ref{f1}) to be different from zero, 
the operator $e^{i\frac{2\pi }{L}\hat{X}}$, when decomposed into irreducible 
representations, should contain the trivial representation of the space 
group. {\em This is not the case if }$n_{0}$ {\em is not an integer}. It is 
straightforward to check that $
\hat{T}\ e^{i\frac{2\pi }{L}\hat{X}}\ \hat{T}^{\dagger }=e^{-i2\pi n_{0}}e^{i% 
\frac{2\pi }{L}\hat{X}}\ .$ 
Then $\langle g|e^{i\frac{2\pi }{L}\hat{X}}|g\rangle =\langle g|\hat{T}% 
^{\dagger }\hat{T}\ e^{i\frac{2\pi }{L}\hat{X}}\ \hat{T}^{\dagger }\hat{T}% 
|g\rangle =e^{-i2\pi n_{0}}\langle g|e^{i\frac{2\pi }{L}\hat{X}}|g\rangle$, 
and the matrix element should vanish unless $n_{0}$ is an integer. This 
restriction is too severe. For example, in 1D a Mott transition 
can take place for any filling, particularly if $n_{0}$ is given by a simple 
fraction \cite{gia}. For non-integer fillings, the definitions of Refs. 
\cite{res1,res2} 
lead to an undefined $\langle \hat{X}\rangle$ and infinite $\lambda$ 
{\em for any} system which is {\it incorrect}. 
 
The correct definition for arbitrary fillings 
$n_{0}=n/l$, where $n/l$ is an irreducible fraction, is  
\begin{equation} 
z_{L}\left[n/l\right]=\langle g|e^{i\frac{2\pi }{L}l\hat{X}}|g\rangle \ . 
\label{f5} 
\end{equation}
Then, the GS expectation value of the position 
operator is defined (mod($L/l$)) as 
\begin{equation} 
\langle \hat{X}\left[ n/l\right] \rangle =\frac{L}{2\pi l}% 
%TCIMACRO{\func{Im} } 
%BeginExpansion 
\mathop{\rm Im}% 
%EndExpansion 
\ln z_{L}\left[ n/l\right] \ ,  \label{f6} 
\end{equation}
and similarly to Ref. \cite{res2} one can introduce a localization length 
$\lambda$. The first important point in these definitions is that the 
operator $e^{i \frac{2\pi }{L}l\hat{X}}$ is invariant under translations and 
has no definite parity under space inversion. This ensures that the matrix 
element entering Eq. (\ref{f5}) is not zero in finite systems, except for 
very particular cases 
in which hidden quantum numbers exist, like for free electrons, in which not 
only the total wave vector $K$, but also the one-particle wave vectors are 
conserved by the Hamiltonian, while the latter are shifted by the operator 
$e^{i\frac{2\pi }{L}l\hat{X}}$ \cite{epl,note4}. The second important remark 
is that, for non-integer $n_{0}$, when a family of Hamiltonians is introduced 
in which the one-particle wave vectors are shifted by a parameter $\alpha$ 
(Eq. (5) of Ref. \cite{res1}), there is a crossing of levels as a function 
of $\alpha $ between states of different $K$'s. Crossings of 
these type were found previously when calculating Berry's phases in systems 
which show the so-called anomalous flux quantization characteristic of 
superconductors \cite{epl,gag}, and are {\it harmless}, since the states 
involved have different quantum numbers and, therefore, are not mixed by the 
Hamiltonian. However, as a consequence of this crossing, Eq. (7) of Ref.  
\cite{res1}, stating that $e^{i\frac{2\pi }{L}\hat{X}}|g\rangle =e^{i\gamma 
_{L}}|g\rangle +{\cal O}(1/L)$ (where $\gamma _{L}$ is a geometric phase) is  
{\it incorrect}, since both members have different $K$'s for non-integer 
$n_{0}$. Finally, the behavior of $z_{L}$ in the thermodynamic limit provides 
us with a universal criterion to distinguish between a conductor (metal or 
superconductor, as we will see below) and an insulator: $z_{L}$ vanishes in 
the first case, while  $|z_{L}|\rightarrow 1$ in the second. 

In the general case (except for the above mentioned particular cases with 
hidden symmetries) one has $e^{i\frac{2\pi }{L}l\hat{X}}| g\rangle = 
e^{i\gamma _{L}}|g\rangle +{\cal O}(1/L)$, and the rest of the demonstration 
leading to Eq. (\ref{f6}) can be done following Ref. \cite{res1} with simple 
and straightforward changes. As shown elsewhere \cite{epl}, the Berry's 
phase $\gamma _{L} = 2 \pi l\langle \hat{X}\rangle /L$ is 
\begin{equation} 
\gamma _{L}=\int_{0}^{2\pi l}d\Phi \ \langle g_{K}(\Phi )| 
\frac{\partial}{\partial \Phi } g_{K}(\Phi ) \rangle \ ,  \label{f7} 
\end{equation} 
where $|g_{K}(\Phi) \rangle$ is the state obtained by adiabatic continuation 
of the GS $|g\rangle$, when a flux $\Phi$ is threaded 
through the ring \cite{Note} (a 1D system with BvK BC is topologically 
equivalent to a ring.) 

Using $z_L$ to distinguish between a conducting and a non-conducting system 
is more stringent than the criterion introduced by Kohn \cite{kohn}. Kohn's 
criterion, universally used in actual calculations, is based on the value of 
the charge stiffness $D_{c}$ (i.e. sensitivity of the curvature of the total 
energy to phase changes in the BC): in the thermodynamic limit
$D_{c}$ is zero in the insulating phase, while 
it can attain any positive value in a conducting system. Instead, 
%it is clear that $|z_{L}| \leq 1$ for any finite system, 
$\lim_{L \rightarrow
\infty} |z_{L}| =0$ for conducting systems, and $\lim_{L \rightarrow 
\infty} |z_{L}| =1$ for non-correlated insulators, or systems in which all 
particles are localized in non overlapping sectors \cite{note4}. To prove 
that this last limit can be achieved for {\it any} correlated 
insulator is rather involved, but one can prove, for a generic insulating 
state, that for large $L$, $|z_{L}| \simeq 1 - D_c/n_0$, and then 
$\lim_{L \rightarrow \infty} |z_{L}| = 1$.
  
To illustrate the main concepts described above, we consider an extended 
Hubbard model (EHM) with arbitrary nearest-neighbor interaction $V$ and 
infinite on-site repulsion $U$ in 1D, at quarter filling ($n_{0}=1/2$). By 
a straightforward extension of the methods used in similar models \cite{cas}, 
it can be shown that this is mapped onto a simple model of 
spinless interacting fermions \cite{note2}, for which some analytical 
results can be obtained in the thermodynamic limit. We explore the 
possibility of quantum phase transitions using our generalizations $z_{L}$ 
and $\langle \hat{X}\rangle$. Finally, we discuss two 
generalizations of this model. 
 
The Hamiltonian which describes the charge dynamics of the model in a ring 
of $L=2N$ sites with BvK BC (${c}_{j+L}^{\dagger }={c}_{j}^{\dagger }$) 
threaded by a flux $\Phi $ is 
\begin{equation} 
H(\Phi)=-t\sum_{j}(e^{i\frac{\Phi }{L}}c_{j+1}^{\dagger}c_{j}^{\;}
+h.c.)+V\sum_{j}n_{j+1}n_{j} .  \label{f8} 
\end{equation} 
The gauge transformation ${\bar{c}}_{j}^{\dagger }=e^{ij\Phi 
/L}c_{j}^{\dagger }$ transforms $H(\Phi )$ into a Hamiltonian $\bar{H}(\Phi 
) $ in which the phase factors disappear at the cost of introducing twisted 
BC ${\bar{c}}_{j+L}^{\dagger }=e^{i\Phi }{\bar{c}}_{j}^{\dagger }$ except 
for fluxes $\Phi =2\pi \times  
%TCIMACRO{\func{integer}} 
%BeginExpansion 
\mathop{\rm integer}% 
%EndExpansion 
$ \cite{note2}. 
 
For $V\gg t$, the GS can be obtained by perturbation 
theory. For $t=0$ it is two-fold degenerate between the 
charge-density-wave (CDW) states $|1\rangle$ and $|2\rangle$, with the 
particles occupying every second site: $|1\rangle = 
\Pi_{j=0}^{N-1}c_{2j+1}^{\dagger }|0\rangle $, $|2\rangle = 
\Pi_{j=1}^{N}c_{2j}^{\dagger }|0\rangle$. These states are mixed in $N^{th}$ 
order perturbation theory by $2N!$ processes in which all particles hop 
either to the left or to the right in some order. For odd $N$ \cite{note}, 
the effective matrix element is $-r(N)(t/V)^{N} \cos (\Phi /2)$, where $r(N)$ 
is a real positive number with $1<r(N)/2<N!$. Thus, the eigenstates are 
$|g_{0}(\Phi)\rangle = (|1\rangle +|2\rangle )/\sqrt{2}$ and $|g_{\pi 
}(\Phi)\rangle = (|1\rangle -|2\rangle )/\sqrt{2}$. The former (latter) is 
the GS for $\Phi=0$ ($\Phi=2\pi$). It is interesting to note what 
happens if $\Phi$ is varied adiabatically from $0$ to $2\pi$: the wave 
vector $K=0$ of the GS of $H(\Phi)$ remains the same, while the 
wave vector $\bar{K}$ in the representation of $\bar{H}(\Phi)$ varies as 
$\bar{K}=K+\Phi n_0$ in general \cite{epl}. Since $\bar{H}(2\pi)=\bar{H}(0)$, 
after completing this cycle, $|g_{0}(\Phi )\rangle $ has evolved from the 
GS of $\bar{H}(0)$ with $\bar{K}=0$ to its first excited state 
with $\bar{K}=\pi$. After another cycle $|g_{0}(4\pi)\rangle$ returns to the 
GS of $\bar{H}(0)$ except for the geometrical phase Eq. (\ref{f7}). 
It is immediate to see that $\hat{X}|2\rangle = 
\sum_{j=1}^{N}(2j)|2\rangle =N(N+1)|2\rangle$, $\hat{X}|1\rangle = 
N(N-1)|1\rangle$. Then, $e^{i\frac{2\pi }{L}\hat{X}}|g_{0}(\Phi) \rangle = 
\pm |g_{\pi}(\Phi)\rangle$, where the sign depends on the parity of 
$(N+1)/2$. This leads to a vanishing $z_{N}$, as defined by Eq. (\ref{f1}), 
but to a $z_{L}[1/2]=1$. 
 
For attractive $V$ and $|V|\gg t$, all particles group together. For $t=0$ the 
degenerate GS is composed of the states $|PS\rangle_j = 
\hat{T}^{j}|PS\rangle$ ($0\leq j\leq N-1$), and $|PS\rangle =\Pi_{j=1}^{N}
c_{j}^{\dagger }|0\rangle$. Then, $e^{i\frac{4\pi}{L}\hat{X}}|PS\rangle_j = 
(-1)^{N+1} |PS\rangle_j$ leading to $z_{L}[1/2]=(-1)^{N+1}$. 
 
For arbitrary values of $V$ and $t$, the spinless model is equivalent (via a 
Jordan-Wigner transformation) to an XXZ model $\sum_{<ij>\alpha }J_{\alpha 
}S_{i}^{\alpha }S_{j}^{\alpha }$ with $N$ spins up, and $J_{x}=J_{y}=2t$, 
$J_{z}=V$. This model was solved by the Bethe ansatz \cite{yan}. From the 
solution one knows that there is a Mott transition at $V=2t$ and a transition 
to the phase segregated (PS) state at $V=-2t$. We have calculated the 
correlation exponent  $K_{\rho }$ of the EHM in the 
conducting phase ($-2t<V<2t$), solving the Bethe-ansatz integral equations 
for the energy per site $e(n_{0})$ of the equivalent XXZ model \cite{yan}, 
and using the expression $K_{\rho }=\pi \sqrt{D_{c}/(2\partial ^{2}e/\partial 
n_{0}^{2})}$, where the Drude weight (or charge stiffness) is given by the 
Bethe ansatz result $D_{c}=2\pi t\sin \mu /[8\mu (\pi -\mu )]$ with $\mu 
=\arccos (V/2t)$ \cite{sri}. We obtain that $K_{\rho }>1$ 
(superconducting correlations dominate at large distances) for $-2t<V<V_{c}$, 
with $V_{c}=-\sqrt{2}t$ within our accuracy ($10^{-3}$).  
It is clear that this model provides a rich zero temperature phase 
diagram and an interesting laboratory to study localization and transitions 
to superconducting (SC) states. 
 
Unfortunately the Bethe-ansatz wave function is quite difficult to handle. 
Therefore, we obtained the GS $|g\rangle$ for finite systems with 
up to $L=16$ sites by the Lanczos method. As usual, $|g\rangle$ was taken 
at the value of $\Phi $ which minimizes the GS energy 
$E_{g}(\Phi)$. In the spinless model, this corresponds to $\Phi=0$ (BvK BC) 
for odd $N$ and $\Phi=\pi$ (antiperiodic BC) for even $N$, and in both cases 
$\bar{K}=0$ \cite{note2}. This choice reduces the dependence of the 
GS energy and $z_{L}$ with size and also leads to a slightly more 
abrupt change in $|z_{L}|$ near the transitions. The resulting $|z_{L}|$ 
calculated with Eq. (\ref{f5}) is represented in Fig. \ref{fig1} for various 
sizes, together with the corresponding numerical results for 
$D_{c}=(L/2)\partial^{2}E_{g}(\Phi)/\partial \Phi^{2}$, and the above 
mentioned Bethe-ansatz result for $D_{c}$ in the thermodynamic limit 
($L\rightarrow \infty $). From the latter quantity, the conductor-insulator 
transitions at $|V|=2t$ are evident. However, the finite-size results always 
lead to a non-zero $D_{c}$ and except perhaps for the change in curvature 
near $V\sim -2.5t$ there are no clear indications of any transition. The 
situation improves if the finite-size results are extrapolated using a 
polynomial in $1/L$. The extrapolation agrees very well with the exact result 
in the conducting phase, and suggests a transition near $V=-2t$. However, 
from the extrapolated $D_{c}$ no conclusions concerning the Mott transition 
at $V=2t$ can be drawn. Only the large size dependence near $V\sim 3t$ is 
indicative of a charge gap.  

Near $V=-2t$, as $V$ increases, $|z_{L}|$ decreases abruptly from values 
near one, to very small values, as expected for an 
insulator-conductor transition. For $V\sim 2t$, the change in  
$|z_{L}|$ is rather smooth and the size dependence is small, but also the 
change of behavior becomes more abrupt with increasing size. The 
extrapolated values of $|z_{L}|$ clearly show an abrupt transition near 
$V=-2t$, indicating that the system is a conductor (very small $|z_{L}|$) for 
$-1.8t<V<1.2t$, and suggesting that it is an insulator for $V\sim 3t$. 
Notice that the convergence to the thermodynamic limit for $D_c$ and $|z_{L}|$ 
is, in absolute value, about the same. However, for any value of 
$V$, $D_{c}$ decreases with system size while, in general, $|z_{L}|$ 
decreases well inside the metallic phases ($|V|<1.5t$) and increases well 
inside the insulating phases ($|V|\sim 3 t$). Thus, the 
results for $|z_{L}|$ are complementary to Kohn's $D_c$ criterion 
\cite{kohn}, $|z_{L}|$ providing a more useful measure. 
 
We have also studied the behavior of $\gamma _{L}$ (Eq. 
(\ref{f7})) near the transitions. Because of inversion symmetry, 
$\gamma _{L}=0$ or $\pi $ (mod($2\pi$)). In contrast to previous cases  
\cite{epl,gag}, we do not find a jump in $\gamma _{L}$ at the transition 
from the SC to the PS regime ($V=-2t$), or at the 
opening of the charge gap ($V=2t$). However, as in the case of attractive $U$ 
\cite{epl}, there is a jump from $\gamma _{L}=\pi$ 
to $\gamma _{L}=0$ as the dominant correlation functions at large distances 
change from the superconducting to the CDW ones. For $8\leq L\leq 16$ we 
obtain with four digits accuracy $V_{c}=-\sqrt{2}t$ in perfect agreement 
with the value obtained from the numerical solution of the Bethe-ansatz 
integral equations. Instead, the corresponding jump in the phase of $z_{L}$ 
takes place for $V_{c}\sim -1.55t$. This suggests that using $\langle 
\hat{X} \rangle =L\gamma _{L}/(2\pi l)$ and Eq. (\ref{f7}) one obtains a 
faster convergence for $\langle \hat{X}\rangle $ to the thermodynamic limit 
than using Eq. (\ref{f6}). In addition, since in the gapless metallic phases 
$\lim_{L\rightarrow \infty }|z_{L}|=0$, it might be difficult to identify 
$V_{c}$ from the corresponding zero of $z_{L}$ for sufficiently large $L$. 
As in Ref. \cite{epl}, we also find jumps in $\gamma _{L}$ for other 
values of $V$, in particular for $V=0$ and $V=t$, without an obvious 
physical meaning. 
 
In the rest of this Letter, we discuss in more detail the expectation value 
of the position operator, Eq. (\ref{f6}), and illustrate its physical meaning 
using two other simple examples. First, note that if $|g\rangle$ in Eq. 
(\ref{f6}) is replaced by one of the CDW states $|1\rangle$ or $|2\rangle$ 
discussed above, the result is $\langle \hat{X}\rangle = 0$ in both cases. 
This is reasonable, since $|1\rangle$ and $|2\rangle$ differ in a 
translation, and are thus physically equivalent in translationally invariant 
systems. This is, in general, the reason why $\langle \hat{X}\rangle $ is 
defined mod($L/l$). Assume now that in the model Eq. (\ref{f8}) one 
considers an attractive $V$ and adds a next-nearest-neighbor repulsion 
$V^{\prime}$ such that $V^{\prime}\gg |V|\gg t$. For even $N=L/2$, 
$|g\rangle$ consists of a sequence of two nearest-neighbor sites occupied, 
the next two empty, and so on. Specifically $|g_{\pm \pi /2}(\Phi)\rangle =
(|A\rangle \pm i\hat{T}|A\rangle )/\sqrt{2}$, with $|A\rangle = 
\Pi_{j=0}^{N/2-1}c_{4j+1}^{\dagger}c_{4j+2}^{\dagger}|0\rangle$. In addition,  
$\hat{X}|A\rangle=\sum_{j=0}^{N/2-1}(8j+3)|A\rangle=N(N-1/2)|A\rangle$. Then, 
$e^{i\frac{4\pi}{L} \hat{X}}|g_{\pm \pi/2}\rangle = -|g_{\pm \pi /2}\rangle$, 
and Eq. (\ref{f6}) gives $\langle \hat{X}\rangle = L/4=N/2$ (mod($L/2$)). 
This is consistent with the transfer of half of the particles of the system, 
either to the left or to the right, necessary to build the state $(|A\rangle 
\pm i\hat{T}|A\rangle)/ \sqrt{2}$ from $(|1\rangle \pm i 
\hat{T}|1\rangle)/\sqrt{2}$. 
 
In the examples discussed above, the Hamiltonian had inversion symmetry.  
Under those circumstances it can be shown that the phase of $z_{L}$ can only 
attain the values $0$ or $\pi$ ($z_{L}$ is real). This statement 
implies that well-defined values of $\langle \hat{X}\rangle$ different from 
0 and $L/(2 l)$ (mod($L/l$)) can only be achieved in the absence of inversion 
symmetry. A simple example in which this symmetry is explicitly broken is 
obtained replacing each site in Eq. (\ref{f8}) by an heteronuclear molecule. 
Specifically, in the spinless case consider a system described by the 
Hamiltonian 
\begin{eqnarray} 
H(0) &=&\Delta \sum_{j}d_{j}^{\dagger}d_{j}^{\;}- 
\sum_{j}(t^{\prime} \ d_{j}^{\dagger}c_{j}^{\;}+ t \ c_{j+1}^{\dagger} 
d_{j}^{\;}+h.c.) \nonumber \\
&&+\sum_{j} (V^{\prime} \ c_{j}^{\dagger} c_{j}^{\;} d_{j}^{\dagger} 
d_{j}^{\;}+V \ c_{j+1}^{\dagger} c_{j+1}^{\;} d_{j}^{\dagger} 
d_{j}^{\;} ) \ .  \label{f9} 
\end{eqnarray} 
It is easy to see that for $V,V^{\prime},t^{\prime}\gg t$, and $n_{0}=1/2$ 
(one particle each two unit cells), $|g\rangle$ is a 
CDW with every second molecule singly occupied in the GS of the 
first two terms of Eq. (\ref{f9}). For this $|g\rangle$, Eq. (\ref{f6}), 
gives $\langle \hat{X}\rangle = NP$ where $P$ is the polarizability of each 
molecule and varies continuously as a function of $t^{\prime}/\Delta$ 
\cite{ort}.  

In conclusion, we have introduced a complex quantity $z_L$ which is shown to 
display a qualitatively different behavior for conductors than for 
insulators, thereby providing a valuable criterion to distinguish between 
those states of 
matter. The criterion, which is valid for any fractional particle filling
and for both, ordered and disordered systems, is 
complementary and in a sense sharper than that based on the value 
of $D_c$ \cite{kohn}. 
We have shown how to use $z_L$ to characterize Mott and 
superconducting transitions in models of strongly correlated quantum 
particles. 
$z_L$ involves {\it only} the computation of the 
GS of the system, which for time-reversal symmetric Hamiltonians is 
real valued, whereas $D_c$ needs at least two GS 
calculations for different BC and in one of them, time-reversal symmetry
is broken. This seems innocuous for Lanczos 
studies, but it is not for stochastic approaches where the study of 
non-time reversal symmetric states adds an additional complication 
to the already infamous fermion sign problem. 
 
One of us (A.A.A.) is partially supported by CONICET, Argentina. G.O. 
acknowledges support from an Oppenheimer fellowship.

\begin{figure}[htb]
\caption{Drude weight $D_c$ (top) and modulus of $z_{L}$ defined by Eq. 
(\ref{f5}) (bottom) as a function of $V/t$ for the $U \rightarrow + \infty$ 
EHM at filling $n_{0}=1/2$ for rings of various lengths 
$L$. Also shown are the corresponding quantities extrapolated to the 
thermodynamic limit using a cubic polynomial in $1/L$, and the exact result 
for $D_c$ in the thermodynamic limit. The vertical lines separate the
regions of phase segregation (PS), Luttinger liquid with $K_{\rho}>1$ (SC) or 
$K_{\rho}<1$ (C) and a new insulating phase at large $V/t$ (I).}
\label{fig1}
\end{figure}
 
\end{multicols}


\begin{references}
\bibitem{note4} $e^{i\frac{2\pi }{L}l\hat{X}}$ shifts all one-particle wave
vectors by $-2 \pi L/l$. Thus, in Fermi or Luttinger liquids, the Fermi surface
is shifted by $e^{i\frac{2\pi }{L}l\hat{X}}$ and 
$z_L$ (Eq. (\ref{f5})) should vanish for $L \rightarrow \infty$. In the 
insulating phases of the models treated here and in Refs. \cite{res1,res2}, 
$|z_L|=1$ for $L \rightarrow \infty$ in the weak and strong-coupling regimes. 
In 
the latter, the particles are localized at different sites (or molecules) and 
$e^{i\frac{2\pi }{L}l\hat{X}}$ reduces to a phase, while in the weak coupling
limit, a Hartree-Fock treatment leads to a symmetry breaking and folding of the
Brillouin zone, and $|z_L|=1$, as in non-correlated insulators.
\bibitem{Note}  The GS $|g_{K}(\Phi )\rangle $ is a section of the 
fibre bundle determined by the twisted BC.
\bibitem{note2}  The mapping from the EHM for 
$U\rightarrow +\infty $ to the spinless fermion model affects the BC in a way 
equivalent to introducing a shift $\Delta \Phi$ in the flux entering 
Eq. (\ref{f8}) which depends on the spin configuration \cite{cas}. This shift 
is irrelevant for our discussion. Note, however, that the twisted BC which 
lead to the minimum GS energy $E_{g}(\Phi +\Delta \Phi)$ in the 
EHM, depends on the spin configuration.
\bibitem{note}  For even $N$ the situation only differs in the GS
total momentum ($K=\pm \pi/2$ depending on $\Phi$), while the energy is 
minimum for $\Phi=\pm \pi$ (antiperiodic BC in $\bar{H}$, with $\bar{K}=0$).
\end{references}
\end{document}